\documentclass[aps,prl,reprint,bibnotes,footinbib,showkeys,groupedaddress,superscriptaddress]{revtex4-1}
\usepackage{graphicx,amsfonts}
\usepackage{amsmath,empheq}
\usepackage{verbatim}
\usepackage{amssymb}
\usepackage{amsthm}

\usepackage{natbib}
\usepackage{epic}
\usepackage{natbib}
\usepackage{xcolor}
\usepackage{color}
\definecolor{darkblue}{rgb}{0,0,0.5}
\definecolor{darkred}{rgb}{0.5,0,0}
\usepackage[colorlinks=true,urlcolor=darkblue,citecolor=darkblue,linkcolor=darkred,hyperfootnotes=false]{hyperref}
\usepackage{textcomp}
\usepackage{cleveref}
\usepackage[normalem]{ulem}

\newcommand{\dive}[1]{\nabla\cdot\mathbf{#1}}

\newcommand{\M}[1]{\mathcal{#1}}
\newcommand{\bo}[1]{\mathbf{#1}}
\newcommand{\T}[1]{\text{#1}}

\newcommand{\h}[1]{\hat{#1}}
\newcommand{\tl}[1]{\tilde{#1}}

\newcommand{\sg}{\text{sgn}}
\newcommand{\fa}{f^{\text{nl}}}

\newcommand{\bos}{\boldsymbol}

\begin{document}

\title{Drift induced by dissipation}

\author{Reinaldo  Garc\'{i}a-Garc\'{i}a}
\email{reinaldomeister@gmail.com}
\affiliation{Laboratoire de Physico-Chimie Th\'eorique-UMR CNRS Gulliver 7083, PSL Research University,
ESPCI, 10 rue Vauquelin, 75231 Paris cedex 05, France}
\affiliation{PMMH, CNRS UMR 7636, PSL Research University, 
 ESPCI, 10 rue de Vauquelin, 75231 Paris cedex 05, France}

\author{Pierre Collet}
\email{Pierre.Collet@cpht.polytechnique.fr}
\affiliation{
Centre de Physique Th\'eorique, CNRS UMR 7644  \'{E}cole Polytechnique, 91128 Palaiseau Cedex, France}
\author{Lev Truskinovsky}
\email{trusk@lms.polytechnique.fr}
\affiliation{PMMH, CNRS UMR 7636, PSL Research University, 
 ESPCI, 10 rue de Vauquelin, 75231 Paris cedex 05, France}

\begin{abstract}	
Active particles have become a subject of intense interest across several disciplines from 
animal behavior  to granular physics. Usually the models of such  particles contain an explicit internal driving.
Here we propose a  model with implicit driving in the sense that the behavior of our particle  is 
 fully dissipative at zero temperature but becomes active  in the presence of seemingly innocent  equilibrium   fluctuations.
 The mechanism of activity is related to the breaking of the gradient structure  in the  chemo-mechanical coupling. We  show that 
 the thermodynamics of such active particles  depends crucially on inertia and cannot be correctly captured in the standard  Smoluchowski limit.  
 To deal with stall conditions, we generalize the definition of Stokes efficiency, assessing the quality of active force generation.
 We propose  a simple  realization of the model in terms of an electric circuit
 capable of turning  fluctuations into a  directed current without an explicit  source of  voltage.
 \end{abstract}

\maketitle

Motile cells, living bacteria, synthetic swimmers  and `walking' grains  are usually modeled  as Active Brownian Particles (ABP)  
\cite{RevModPhys.85.1143,SR-statphys,romanczuk2012active,RevModPhys.88.045006}. While it is clear that 
to achieve persistence, ABPs need to violate fluctuation dissipation theorem and extract  energy from the environment, the  underlying  
mechanisms of time  reversal symmetry (TRS)  breaking at 
the microscale  are known only in few cases \cite{reimann2002brownian, PhysRevX.7.021007}.  
Moreover,  even in those cases,  the stochastic thermodynamics of macroscopic directional drift  is still replete with `hidden  effects'  and `anomalies'  
 \cite{Celani-anomaly, PhysRevE.88.022147, PhysRevLett.119.258001, shankar2018hidden}.  
Directionality is usually imposed   through the
asymmetry of the background potential or the explicit external  gradients
 \cite{bodeker2010quantitative,friedrich2007chemotaxis,selmeczi2008cell,
kareiva1983analyzing,komin2004random,howse2007self,paxton2004catalytic}, however,  
it can  also  arise   from   velocity-dependent forces \cite{Schweitzer1998,Badoual-bidirectional, Sarracino1,0295-5075-62-2-196, Sarracino2,Vicsek201271,RevModPhys.85.1143} 
allowing the effective friction
coefficient to be  negative~\cite{romanczuk2012active, ganguly2013stochastic, chaudhuri2014active}.
Such forces are then capable of `pushing' 
the particle and their activity  
can be  interpreted as the presence of  `anti-dissipation' at the microscale.

In this Letter we study a more subtle  mechanism of directional motility  which  relies  on velocity dependent forces  with strictly  positive effective viscosity coefficient.   
Consider, for instance,  an inertial dynamics of  a particle  $ m \dot{v}  =F+f$,  where $m$ is the mass of the particle, $f$ is an external fixed load,
$ F= - \h \gamma(v) v$ is a frictional force and  $\h \gamma  \geq 0$ is an 
 effective friction  coefficient. At zero temperature this system is  clearly  dissipative with  $f v \geq 0$. However,   if one exposes the same particle to an equilibrium  thermal reservoir  
writing dynamics in the form 
 \begin{equation} \label{chemo-mec-dyn0}
 m \dot{v}  =F +f+\xi,
 \end{equation}
where $\langle\xi \rangle=0$   and $\langle \xi(t)\xi (t')\rangle\sim  \delta(t-t')$,  it  may, for particular choices of the function $\h \gamma(v)$,   exhibit `anti-dissipative'  
behavior with  $ f \langle  v \rangle \leq 0$.  In particular,  such particle can behave as a 
Brownian motor  with a nonzero drift  $\langle v \rangle$  at zero $f$, apparently induced by  dissipation. In this Letter we  link   this  
phenomenon with    \emph{non-potential}  structure of dissipation  and \emph{strong} violation of detailed balance (DB).  
We  address the nontrivial nature  of the overdamped  
limit in such systems and show that  the conventional Smoluchowski-type asymptotics  fails to describe adequately the underlying energetics. 
To assess the efficiency of the new motor in the whole range of parameters, we had  to go beyond the  conventional definitions and view stall conditions as a  regime with functional energy consumption. 

To motivate the  model we  consider an underdamped Brownian particle moving in a fluid under the action of viscous friction and thermal noise. 
Suppose that the translational dynamics of the particle  is additionally coupled to a
chemical reaction: 
\begin{empheq}[right=\empheqrbrace] {align}
 \label{chemo-mec-dyn}
 m\dot{\bo v} &=\bo F(\bo v,\Delta\mu)+\bo f+\bos\xi\;\nonumber\\
 \dot\zeta &=A(\bo v,\Delta\mu) 
\end{empheq}
where $\langle\xi_i(t)\rangle=0$, and $\langle\xi_i(t)\xi_j(t')\rangle=2\gamma T\delta_{ij}\delta(t-t')$,
$\gamma$ is the corresponding `bare' viscous coefficient,   $T$ is the temperature of the bath (we set Boltzmann constant equal to one) 
and $A$ is the   driving force acting on the  reaction coordinate $\zeta$. If chemistry and mechanics are decoupled and the particle is in equilibrium, 
we have $\bo F  =-\gamma \bo v$ and  $A  =\Delta \mu\ $ where $\Delta \mu$ is the  affinity of the chemical reaction. To break the TRS we assume  that 
the  fluxes  are related to  forces   through pseudo-Onsagerian relations \cite{RevModPhys.69.1269}

\begin{empheq} [right=\empheqrbrace]{align}
 \label{chemo-mec-dyn1}
\bo F &=-\gamma \bo v +\lambda \Delta \mu (\bo v / \Vert \bo v\Vert ) \;\nonumber\\
A &=  -\lambda (\bo v \cdot \bo m) +  \Delta \mu\;
\end{empheq}
where  the coefficient  $\lambda$ characterizes chemo-mechanical coupling.  The unit vector $\bo m$ indicates  a preferred direction associated, for instance,  
with  an external  concentration gradient. Note that in (\ref{chemo-mec-dyn}, \ref{chemo-mec-dyn1}) the chemical subsystem acts as a feedback controller for  
the mechanical degrees of freedom. 
 
To ensure analytical transparency,  we assume  complete separation of time scales in the sense that
the reaction is stationary   $\dot \zeta=0$. Then   
$
\bo F=-\gamma\big[1-\epsilon\,(\bo m\cdot\bo v)/(\Vert \bo v\Vert)\big] \bo v, 
$
 where  $\epsilon=\lambda^2/\gamma$ is a nondimensional parameter. This form of the friction force  highlights the  underlying anisotropy in dissipation.   
A helpful biological reference for this scenario is bacterial
flagella whose efficiency for self-propulsion crucially depends on the fact that tangential and normal resistance coefficients are different~\cite{0034-4885-72-9-096601}.

Assume  now  that the vector field $\bo m$ is  constant and homogeneous and let us limit our analysis to one dimension.  Then we recover our scalar model \eqref{chemo-mec-dyn0} with  
$
 \h \gamma(v)  =\gamma [1+\epsilon\,\sg(v)], 
$
where $\sg(x)$ is the sign function, see also \cite{Gidoni201465}.   Note that $\h \gamma(v)\neq \h \gamma(-v)$ and therefore this model violates TRS  even when  the system is purely dissipative (for $|\epsilon|<1$). 
If we write   $F(v)=-\gamma v+g(v)$,
where $g(v)$ is  the non-linear  contribution to friction, a broken TRS  
implies that $g(v)\neq-g(-v)$, which is, for instance,  in stark contrast with   the  paradigmatic  
Rayleigh-Helm\-holtz  model  of ABP where always  $g(v)=-g(-v)$ ~\cite{Badoual-bidirectional}
 \begin{figure}[t]
  \includegraphics[scale=0.36]{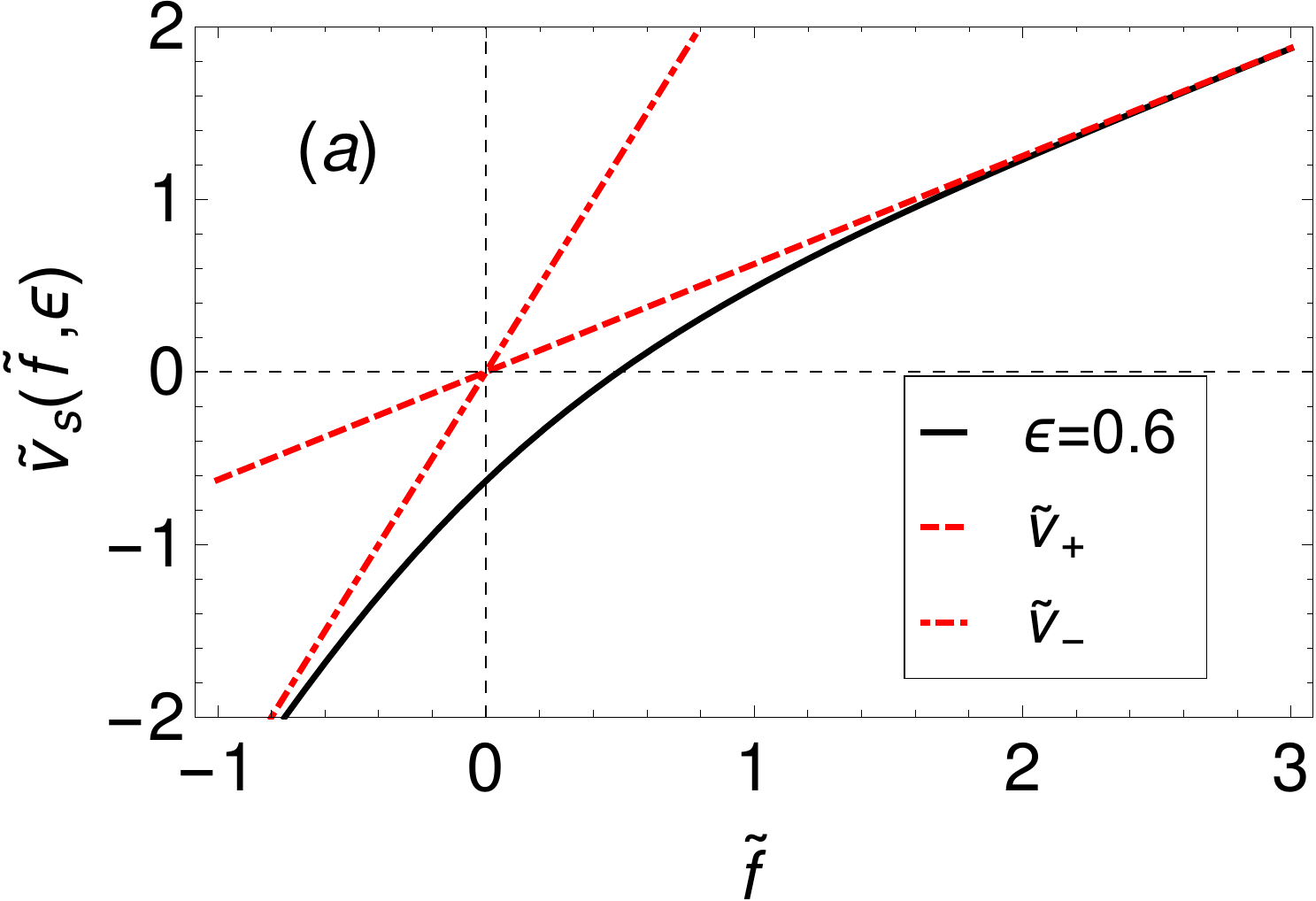}
  \includegraphics[scale=0.35]{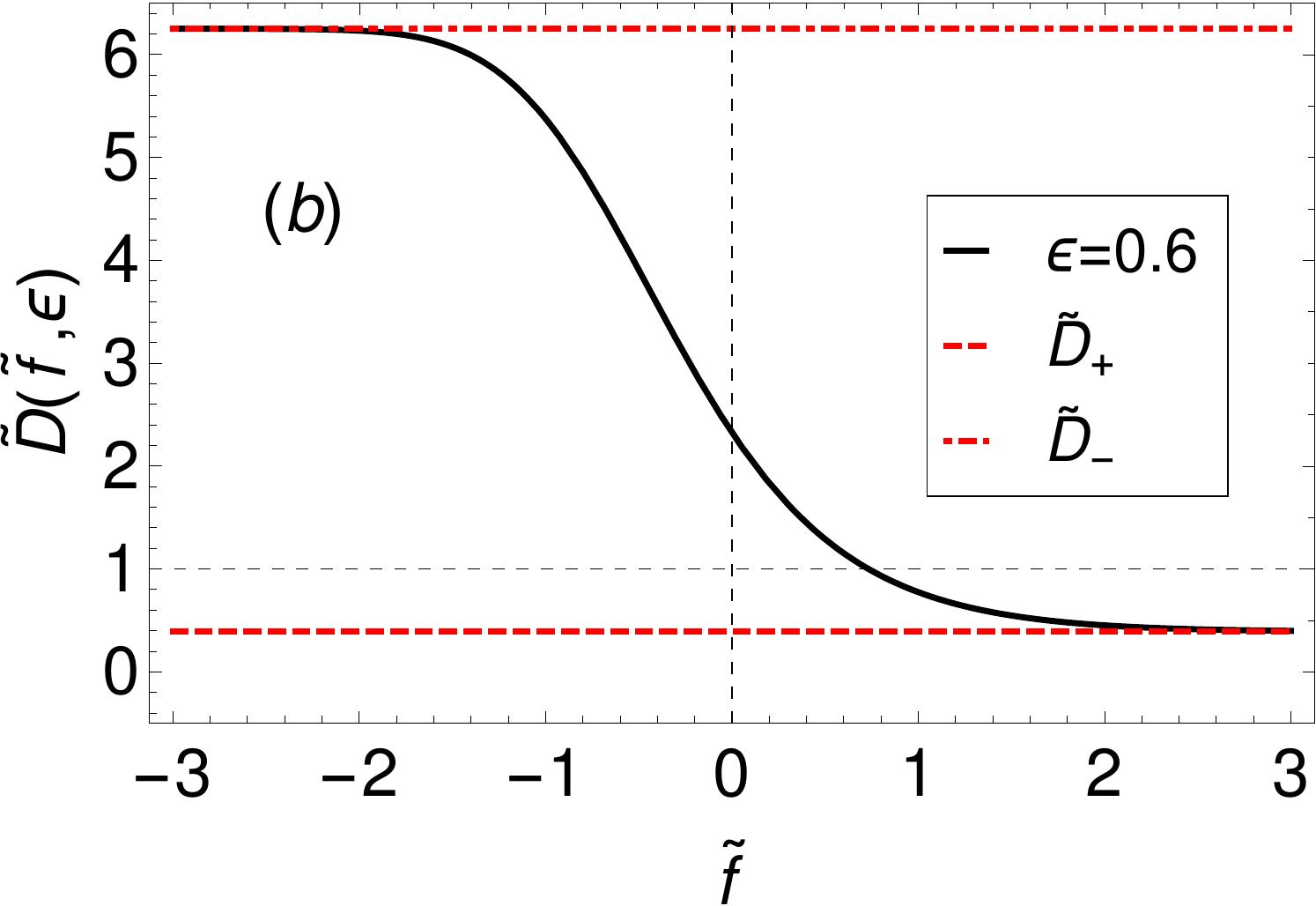}
  \caption{ Stationary response of the system \eqref{detailed-balance1} at $\epsilon=0.6$:  (a) drift velocity  (b) diffusion coefficient. }
  \label{fig1}
\end{figure}
 
To clarify the difference between these two classes of models, assume that 
$F=-\gamma v+\varepsilon g(v)$,  where   
 $g(v)$ is arbitrary and $\varepsilon>0$ is a small parameter. Assume  for generality that the system is also perturbed spatially so that   
 $ m \dot{v}  =-\gamma v+\varepsilon (g(v) + U'(x))+f+\xi$, where $v=\dot{x}$.
 
We begin by writing the  
Kramers equation  for this system, $\partial_t\,P =-\dive{J}$, where  
where  $\bo q=(x,v)$, $P(\bo q,t)$ is  the  probability density   and $\bo J=(J_x, J_v)$ is the probability current   which can be split in a reversible and a 
dissipative parts, $\bo J=\bo J_{\T r}+\bo J_{\T d}$~\cite{Risken1989}, with $\bo J_{\T r}=(v\,P,\varepsilon\,m^{-1}[g_e(v)-U'(x)]P)$, 
and $\bo J_{\T d}=(0,m^{-1}[\varepsilon g_o(v)-\gamma v]P-\gamma Tm^{-2}\partial_v P)$~\cite{Kwon2016,Chaudhuri2016}.
Here we  distinguished between  the \emph{even} and the \emph{odd}  contributions to  the nonlinear force by defining  $g_{e,o}=(g(v)\pm g(-v))/2$.

For the DB condition to be satisfied, we must have  $\bo J_{\T d}=0$, which means that 
$\partial_v \ln P_s=(m/\gamma T)[\varepsilon g_o(v)-\gamma v]$, where $P_s(v,x)$ is the stationary distribution. This
implies that $P_s$ must factorize into the product of a velocity-dependent  and 
 position-dependent functions. In the stationary state we must also have $\dive{\bo J}_{\T r}=0$ or
\begin{equation}
 \label{detailed-balance}
 \partial_x\ln P_s-\frac{\varepsilon}{T}U'=\frac{\varepsilon}{T}\bigg[g_e-\frac{T\partial_v g_e}{m\,v}\bigg]+\frac{\varepsilon^2 g_o}{\gamma\,v T}(g_e-U').
\end{equation}

Since the r.h.s. of \eqref{detailed-balance} cannot  depend on $v$ due to the factorization mentioned above, one must have $g_{e,o}=0$. Moreover,  we see from  \eqref{detailed-balance} that for 
systems with $g_e=0$ but $g_o\neq0$ (Rayleigh-Helm\-holtz  model),  the DB condition holds to first order and  breaks only at   $O(\varepsilon^2)$ (i.e., only in presence of a coupling
with an external potential \cite{Sarracino1,Sarracino2}).  Instead, when  $g_e\neq0$ but $g_o=0$, the DB breaks already at the  first order in $ \varepsilon$ and 
without a need for external interactions. It is then clear  that the degree of non-equilibrium in   systems with  $g_e\neq0$  is fundamentally stronger than in   systems with $g_o\neq0$.
 
To illustrate the behavior of a system with $g_e\neq0$ we  make the simplest assumption  $g(v)  = \epsilon \gamma  \sg(v)\,v$, which implies, in particular,  
that  $g_o=0$ .  We can then drop the irrelevant  potential $U(x)$ and,  
  using dimensionless variables   
$\tilde{v}= v\sqrt{m/T}$, $\tilde{t}= t\gamma/m$, and $\tilde{f}= (f/\gamma)\sqrt{m/T}$,   write the  dynamic equation in the form 
\begin{equation}
 \label{detailed-balance1}
 \dot{\tilde{v}}=-[1+\epsilon\sg(\tilde{v})] \tilde{v}+\tilde{f}+\tilde{\xi},
\end{equation} 
 where  now $\langle\tl \xi(\tl t)\tl \xi(\tl t')\rangle=2\delta(\tl t-\tl t')$. 
  
 The $\tl f$ dependence of  
the steady-state drift velocity $\tilde{v}_s=\langle \tilde{v}\rangle$ can be written explicitly ~\footnote{See Supplemental Material at [URL will be inserted by publisher]} 
and the typical $\tilde{v}_s(f)$ curve, at  $0<\epsilon<1$,  is shown in Fig.\ref{fig1}a.  
In addition to  two  purely  dissipative regimes $\tilde{v}_\pm(f)=\tilde{f}/(1\pm\epsilon)$  reached at $\tilde{f} \rightarrow \pm \infty$  the system also exhibits   
'anti-dissipative'  behavior at small  forces when the particle can carry cargo (to the left, as long as  $\epsilon>0$). A  simple  expression can be obtained  for $f=0$ where the velocity of 
active drift takes its maximum value
\begin{equation}
 \label{drift-zero-force}
 \tilde{v}_s^{\text{m}}(\epsilon)=\sqrt{\frac{2}{\pi}}\frac{\sqrt{1-\epsilon}-\sqrt{1+\epsilon}}{\sqrt{1-\epsilon^2}}<0.
\end{equation}
At $\epsilon\rightarrow0$, we obtain  $-\tilde{v}_s^{\T{m}}\sim \epsilon$  or,  in dimensional variables,      $ -v_s^{\T{m}}\sim \epsilon\sqrt{ T/m}$.  In the presence of  cargo,  the   
same  scaling  can be shown for the \emph{active}  part of the drift  $ v_s^{\text{a}} =v_s -f/\gamma$,  so that again $-v_s^{\text{a}}\sim\epsilon\sqrt{T/m}$ for small $\epsilon$. 
This is a hint  that in  the overdamped regime the  active behavior emerges only in the   limit  when $\epsilon\sim\sqrt{m}$.

The simplicity of the model allows one also to semi-analytically compute the force dependent effective diffusion coefficient  
$
 \tilde{D}=\lim_{\tl t\rightarrow\infty}\langle \tilde{x}^2(\tl t)\rangle-\langle \tilde{x}(\tl t)\rangle^2/(2\tl t), 
$ 
 see  \cite{Note1}  and Fig. \ref{fig1}. The purely dissipative,  large force limits  are again different $\tilde{D}_{\pm}=1/(1\pm\epsilon)^2$,  because the limiting systems  can be  
viewed as  equilibrated with  reservoirs  having different temperatures $\tilde{T}_{\pm}=1/(1\pm\epsilon)$, and viscosities $\tl \gamma_{\pm}=1\pm\epsilon$, so that
$\tl \gamma_+\tilde{T}_+=\tl \gamma_-\tilde{T}_-\equiv1$ and $\tilde{D}_{\pm}=\tilde{T}_{\pm}/\tilde{\gamma}_{\pm} $.  
This observation suggests rewriting our evolution equation  \eqref{detailed-balance1} in the form
\begin{equation}
 \label{equivalent}
 \dot{\tilde{v}} = \begin{cases}
            -\tl \gamma_{-} \tilde{v}+\tilde{f}+\sqrt{2\tl \gamma_-\tilde{T}_-}\xi_-\;\;\;  \tilde{v}<0,\\
            -\tl \gamma_{+} \tilde{v}+\tilde{f}+\sqrt{2\tl \gamma_+\tilde{T}_+}\xi_+\;\;\;  \tilde{v}>0,
           \end{cases}
\end{equation}
with $\langle\xi_i(\tl t)\xi_j(\tl t')\rangle=\delta_{ij}\delta(\tl t-\tl t')$ and  $i,j=\pm$.  This stresses the fact that the activity in this system can be interpreted by the  
exposure of the particle to two reservoirs with different temperatures $\tilde{T}_{\pm}$. Furthermore, such a representation in terms of two Ornstein-Uhlenbeck processes makes
explicit the fact that there are two intrinsic inertial time scales, $\tau_\pm=m \gamma^{-1}_{\pm}$ (in dimensionful units).

\begin{figure}[t]
  \includegraphics[scale=0.37]{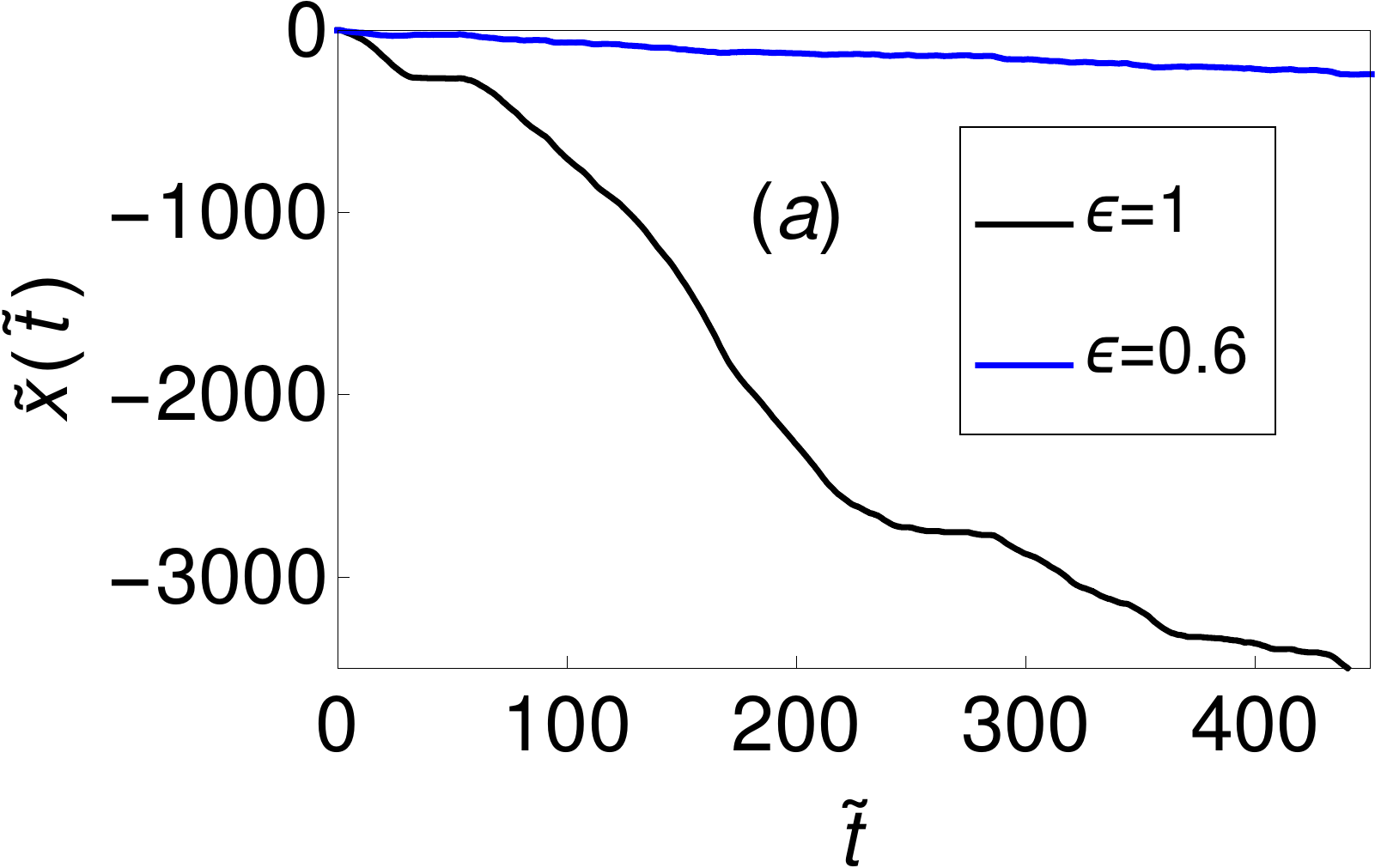}
  \includegraphics[scale=0.35]{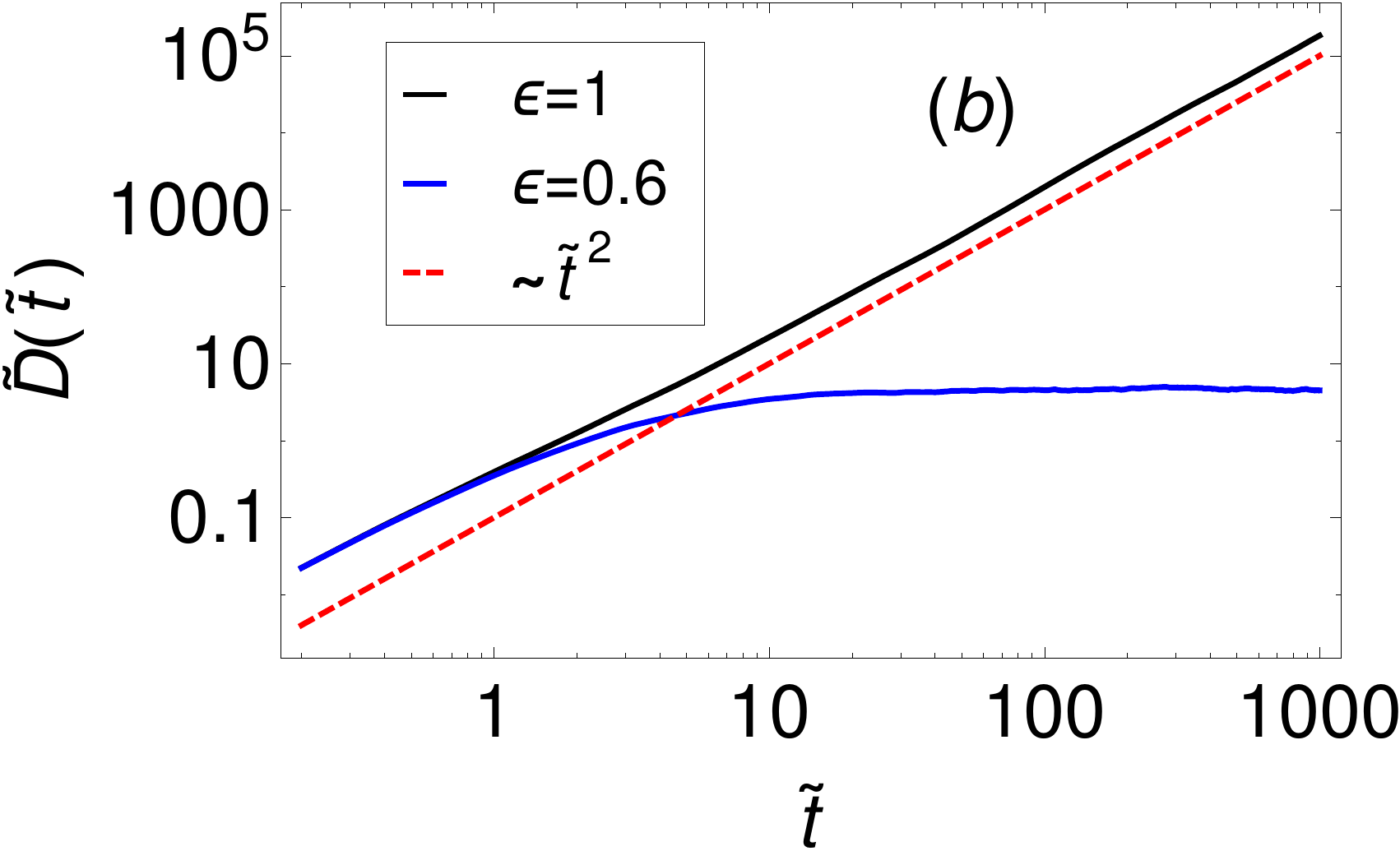}
  \caption{Critical dynamics of a free active  particle at $\epsilon=1$: (a) generated stochastic 
  trajectory (black solid line) versus a stochastic trajectory in the case $\epsilon=0.6$; (b)  finite time diffusion coefficient.}
  \label{fig2}
\end{figure}

Note that at $\epsilon\to1$ the temperature of the `hot reservoir' $\tl{T}_-$  diverges and the velocity dynamics becomes Brownian for $\tilde{v}<0$. As a result 
both the  average  drift velocity  and the recrossing time  (from negative to positive velocity)  
diverge and the dynamics becomes  critical   exhibiting  anomalous unidirectional persistence.  At large times  
one can expect excursions into the preferred  direction to dominate implying
that $\langle\tl x\rangle\sim-\tl{t}^{3/2}$, and $\tl D\sim \tl{t}^2$; the associated transients are illustrated  in Fig.~\ref{fig2}.

Observe next  that in the  double limit $\epsilon \to 0, m \to 0$,  with $\epsilon=-v_s^{*}\sqrt{\pi m/2T}$, when the active drift velocity has a finite limit, $v_s^{\text{a}}\to v_s^*$,   
the (dimensional)  active diffusion coefficient  
$D^a= D-T/\gamma $  disappears with the scaling $ D^a \sim \epsilon T/\gamma$. The limiting  overdamped dynamics, rigorously justified in \cite{Note1},  
takes the form 
\begin{equation}
 \label{overdamped1}
 \dot{x}=\frac{f}{\gamma}+v_s^*+\sqrt{\frac{2T}{\gamma}}\xi.
\end{equation}
which is  often postulated in  phenomenological models of  ABP, e.g.  \cite{Golubeva2012}. Note, however, that the effective model (\ref{overdamped1}), 
being only a weak limit of the original model \eqref{detailed-balance1},  only reproduces trajectories faithfully while misrepresenting the structure of  
velocity  fluctuations which are of order $\sim\sqrt{T/m}$ by equipartition.  
This leads to the appearance of the    `hidden'  terms in the stochastic thermodynamics of such systems, e.g.  \cite{MurashitaEsposito, shankar2018hidden}.

To elucidate this issue we now reintroduce dimensional variables and consider the energetics of a  slightly more general model than \eqref{detailed-balance1}:
\begin{equation}
\label{reduced model}
m\dot{v}=-\gamma v+g_e +f+\xi,
\end{equation}
where  the even function $g_e(v)$ is arbitrary. The energy balance along a particular trajectory of duration $\tau$ can be derived by multiplying~\eqref{reduced model} by $v$ and integrating over time. It reads 
$ \M{E}_\tau=\M U_{\tau}^a-\M{W}_\tau-\M{Q}_\tau$. Here    
 $ \M{E}_\tau=(m/2)[v^2(\tau)-v^2(0)]$ is the change in kinetic energy of the particle,
$\M U_{\tau}^a=\int_0^\tau dtv g_e(v)$ is the  active work performed on the particle, $\M{W}_\tau=-f\int_0^\tau dt v$ is the work  against the load,
and $\M{Q}_\tau=\int_0^\tau dtv(\gamma v-\xi)$ is
the released heat  \cite{sekimoto2010stochastic}. The  stochastic entropy production  can be split into a part associated with the system (particle)  and another part associated with  the reservoir:
$  \M{S}_\tau= \M{S}_\tau^s+ \M{S}_\tau^r$ \cite{Seifert-Review,Note1}.  Here $ \M{S}_\tau^s=\ln [ \rho_0(v(0))/\rho_{\tau}(v(\tau))]$ is the change of the stochastic
Shannon entropy of the  particle, whose velocity at time $\tau$ is distributed with the  probability density $\rho_{\tau}(v)$, and 
$ \M{S}_\tau^r =\M Q_\tau/T- \M{S}_\tau^a$. The quantity
$ \M{S}_\tau^a=m^{-1}\int_0^\tau dt \partial_v\fa_e(v)$ can be interpreted as a leftover of the information exchange  
between the system and the controller after eliminating the controller degrees of 
freedom~\cite{pumping,*Vel-feedback,Munakata1,*Munakata2,*Munakata3,Horowitz-info,  chaudhuri2014active,Chaudhuri2016,Puglisi-1,Puglisi-2,PhysRevLett.119.258001, LeePRL,Kwon2016,PhysRevE.92.032143}. 
To compute the total entropy production we   used the  standard representation $ \M{S}_\tau=\ln(\M P[v]/\h{\M P}[\h v])$~\cite{Seifert-Review}, where $\M P[v]$ is the path 
probability of the trajectory $v$,  while $\h{\M P}[\h v]$ is the  probability for the time-reversed  trajectory $\h v$ \footnote{ In contrast to Refs.~\cite{Puglisi-1,Puglisi-2,PhysRevLett.119.258001}, 
where activity was  modeled by colored noise, and to feedback cooling systems, e.g., Refs.~\cite{pumping,*Vel-feedback,Munakata1,*Munakata2,*Munakata3}, here  we do not need to rely on 
special assumptions about the  time-reversed dynamics, see  ~\cite{Note1} for details.}.

The conventional forms of the first and second laws of thermodynamics can be obtained if we  average the above expressions over the ensemble of possible trajectories and take time derivatives.  
Denote by italic capital letters such averages  and assume that the system is in  a stationary state with $\dot E=0$ and $\dot{S}^s=0$, where for 
instance $\dot{E} =(d/d\tau)\langle \M {E}_\tau \rangle$.  Then we can write: 
\begin{equation}
\label{reduced model2}
  \dot{U}^\text{a}-\dot W-\dot Q=0, \, \,\, \,
 \dot{S}=\dot Q/T-\dot{S}^{\text{a}}=\frac{m^2}{\gamma T}\int\frac{J_d^2(v)}{\rho_s(v)}dv \ge0,
\end{equation}
where $J_d(v)=-m^{-1}[\gamma v+(\gamma T\,m^{-1})\partial_v]\rho_s(v)$ denotes, as before, the dissipative part of the stationary current~\cite{Note1}.

The main shortcoming of the limiting model \eqref{overdamped1} is that it   underestimates entropy production.
Indeed, for the overdamped dynamics \eqref{overdamped1}, the stationary entropy production rate
can be  written as
\begin{equation}
 \label{ent-overdamped}
 \dot{S}_{\text{od}}=\frac{1}{\gamma T}(f+\gamma v_s^{*})^2\geq0.
\end{equation}
It is clearly  associated with passive dissipation described in \eqref{reduced model2} by the term $\dot Q/T$. 
In stall conditions  this expression   vanishes  because  \eqref{overdamped1} 
does not see the  fast dynamics  at the microscale.
If we now compute the  entropy production   for  the full model \eqref{detailed-balance1} and go to the limit 
 $m\to0$ with $ \epsilon\sim\sqrt {m}$ we obtain  \cite{Note1}
\begin{equation}
 \label{missing}
 \dot S = \dot{S}_{\text{od}}+\bigg(\frac{\pi}{2}-1\bigg)\frac{\gamma (v_s^{*})^2}{T}\geq0,
\end{equation}
where the second term  constitutes the `hidden' entropy production.

To assess the efficiency of our ABP it is   natural to first introduce the   injection rate of the  Helmholtz free energy  $\dot{F}^{\text{a}}=\dot{U}^\text{a}-T\dot{S}^{\text{a}}$. 
Then the inequality  in \eqref{reduced model2} can be rewritten as   
$
 \label{first-law1}
 T\dot{S}= \dot{F}^{\text{a}}-\dot{W}\ge0,
$ 
which suggests the  following definition of the  thermodynamic  efficiency,  
$\eta_T=\dot{W}/\dot{F}^{\text{a}}\equiv\dot{W}/(T\dot{S}+\dot{W})\le1$  \cite{Cao-Feito-Feedback,Recho2014}. 
This definition, however, 
neither  accounts for the capacity of a motor to self-propel at zero force, nor for its
ability to generate force in stall conditions: in both limits the   machine works  
(either by achieving   persistent unidirectional displacement or  equally persistent  localization) with apparently 
zero efficiency. A known way to resolve the first of these issues  is to consider the 
Stokes efficiency \cite{0295-5075-57-1-134}, 
$\eta_S=(\dot{W}+\gamma v_s^2)/\dot{F}^{\text{a}}$, which still vanishes in stall  conditions.
\begin{figure}[t]
  \includegraphics[scale=0.35]{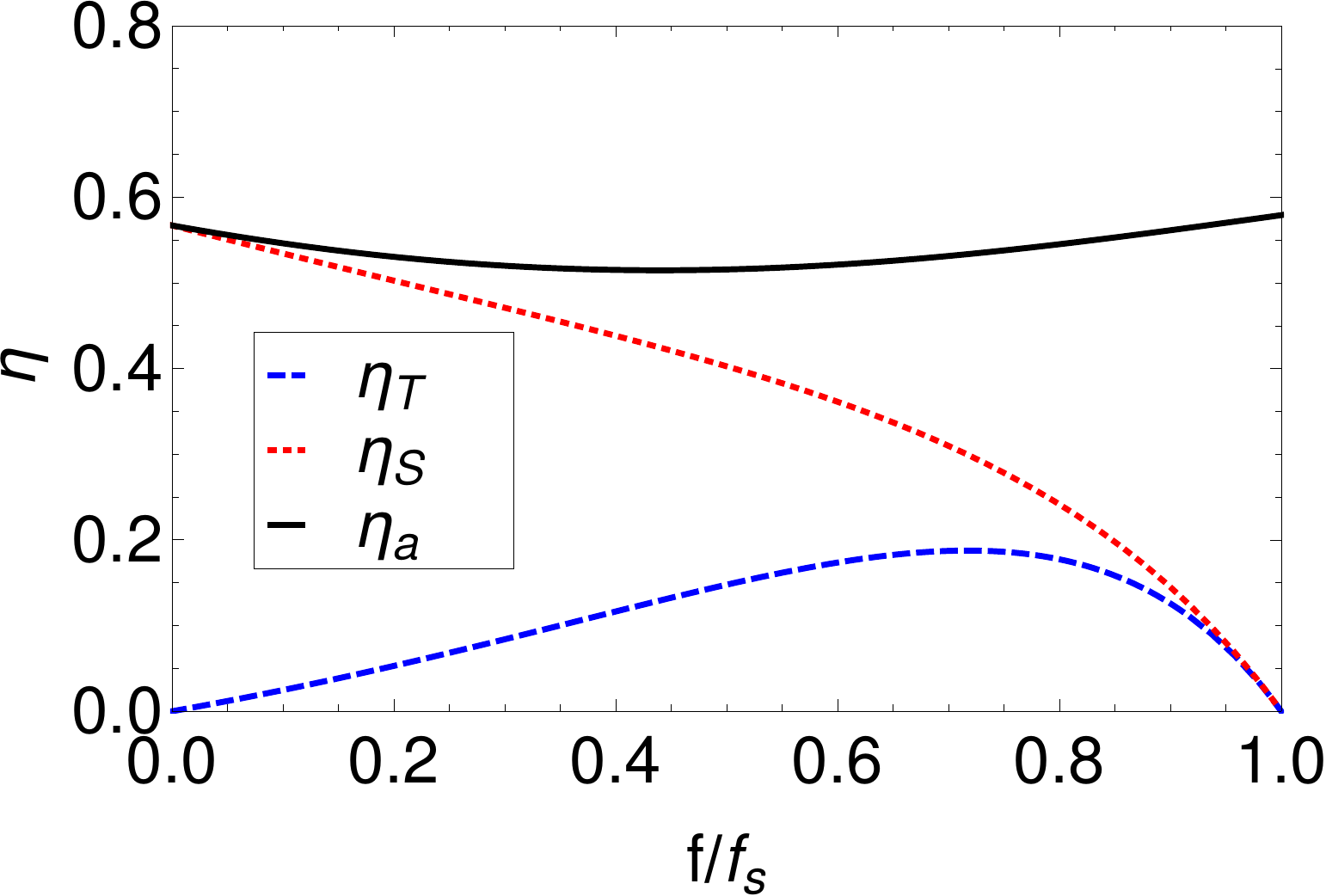}
  \caption{ Typical force dependence of the thermodynamic efficiency $\eta_T$, Stokes efficiency $\eta_S$ and  the new  efficiency  $\eta_{\text{a}}$ for the model~\eqref {detailed-balance1} 
  with   $\epsilon=0.9$. The force $f$ is normalized by the stall value (denoted by $f_s$). }
  \label{fig3}
\end{figure}

To fix this problem we observe that  the (squared) total  active force generated by the controller is $\langle g_e(v)^2\rangle$, while only an amount $\langle g_e(v)\rangle^2$
is useful. The efficiency of active force generation can then be quantified as $\eta_{\text{a}}=\langle g_e(v)\rangle^2/\langle g_e(v)^2\rangle$
 or in  thermodynamic terms~\cite{Note1}
\begin{equation}
\label{new-eff}
\eta_\text{a} = \dot{W}^{\text{a}}/\dot{G}^{\text{a}}.
\end{equation}
Here $\dot{W}^{\text{a}}=\gamma (v_s^{\text{a}})^2$ is now interpreted as useful power, where  $v_s^{\text{a}}=v_s-f/\gamma$ is the  active velocity gain  introduced previously; 
observe that it is finite in both zero  force and zero velocity  limits. The consumed power 
is  naturally measured by the rate of injected Gibbs free energy  $\dot{G}^{\text{a}}=\dot{F}^{\text{a}}-fv_s^{\text{a}} \geq \dot{F}^{\text{a}}$  which is also natural 
given that    the system is performing  work against the load. 
The typical behavior of thermodynamic, Stokes, and force generation  \eqref{new-eff}  efficiencies is illustrated for our model in Fig.~\ref{fig3}.
 Note that the definition \eqref{new-eff} is different from the recently introduced  notion of chemical efficiency~\cite{baerts2017tension} which may attain 
 negative values and does not reduce to  the  Stokes efficiency $\eta_S$  in the absence of load. 

\begin{figure}[t]
 \includegraphics[scale=0.3]{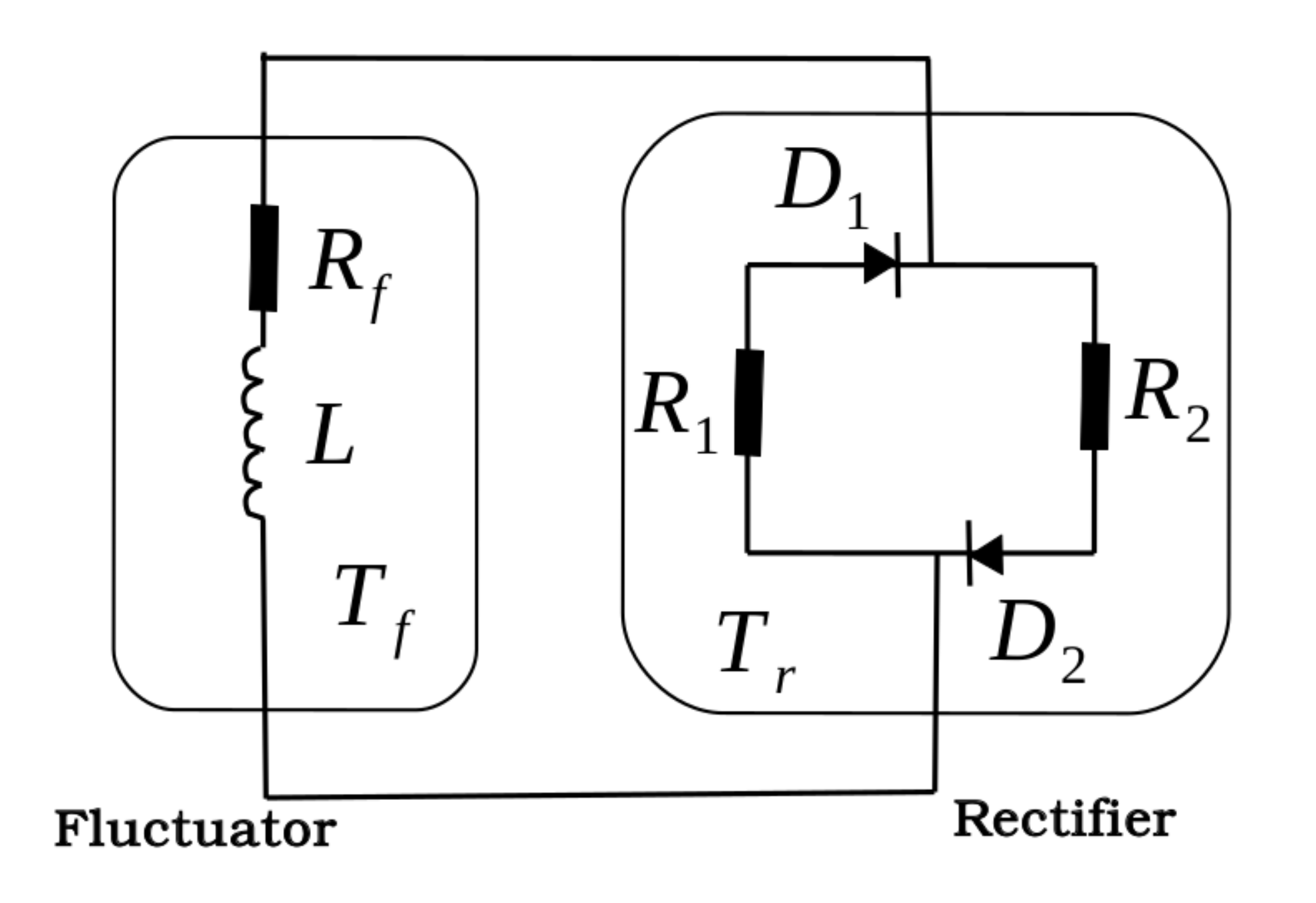}
 \caption{Electric circuit imitating the behavior of the system \eqref{detailed-balance1} when $T_r<<T_f$.}
  \label{fig4} 
\end{figure}

We now briefly discuss a simple experimentally testable  realization of the  system  \eqref{detailed-balance1}
in the form of an electric circuit where  an active current may appear in the absence of directed voltage, see  Fig.~\ref{fig4}.
The `fluctuator' part of the circuit contains an electric resistance $R_f$ and an ideal inductance $L$. We assume  the `fluctuator'  to be  in thermal contact with a bath at temperature $T_f$. 
The `rectifier' is made of two parallel branches, each  containing a resistor $R_i$ and an ideal diode $D_i$ in series ($i=1,2$), and is     
in thermal contact with another bath with  temperature $T_r<<T_f$.  This inequality is essential  to ensure
that electrical fluctuations are  basically produced
only in resistor $R_f$ while the `rectifier'  plays the role of the  active mechanism alternating the  effective resistance depending on the direction   
of the current  in the `fluctuator'.

Standard  circuit analysis  leads to the following equation for the  global current $I$  ~\cite{Note1}:
\begin{equation}
 \label{circuit-analog}
 L\dot{I}= -R_e(1+\epsilon\sg(I))I+\sqrt{2R_e T_e}\xi,
\end{equation}
where  the effective parameters are  $R_e =R_f+(R_1+R_2)/2$, $ T_e  = 2 R_f T_f/(2R_f + R_1+R_2)$ and  $\epsilon  =(R_2-R_1)/(2R_f + R_1+R_2)$. We assumed that the current is positive when
flowing clockwise around the circuit. As the analogy between  \eqref{detailed-balance1} in the absence of load,  and  \eqref{circuit-analog} is complete, 
the circuit in Fig.~\ref{fig4} should be able to generate a directed current by rectifying thermal 
fluctuations; the  `activity' is then ensured by the device  maintaining the temperature difference between the `fluctuator' and the `rectifier'.

To conclude, we presented a  model of an active particle exploiting  'strong' mechanism of TRS breaking. This model appears naturally if one 
makes the simplest pseudo-linear assumptions about the chemo-mechanical coupling of the vectorial (friction) 
and the scalar (reaction) processes which breaks the potentiality of the dissipative potential. While the   realistic chemomecanical coupling is probably   
more complex, for instance quadratic, as in the case of KPZ equation~\cite{KPZ} or the active model B~\cite{Cates-Nat-Commun}, 
 the main idea of the  breaking the TRS symmetry through non-gradient dissipation can be already captured by our semi-analytical model. 
An important result of our analysis is that  in systems with non-Maxwellian velocity distribution and persistence, the Smoluchowski  limit aimed at capturing  
trajectories  can grossly underestimate the associated dissipation, giving a misleading picture of the fluctuation rectification process.

\emph{Acknowledgments.} 
The authors thank J.F. Joanny for helpful discussions. 
R.G.G. acknowledges financial support from the Grant  
ANR-10-LBX-0038 which is a part of the IDEX PSL ( ANR-10-IDEX-0001-02 PSL). 
L. T. was supported by the  Grants ANR-10-IDEX-0 0 01-02 PSL and ANR-17-CE08-0 047-02.

\end{document}